\documentclass[twocolumn]{aastex62}

\submitjournal{ApJ}

\shorttitle{Kinetic models of tangential discontinuities in the solar wind}
\shortauthors{Neukirch et al.}

\begin{document}

\title{Kinetic models of tangential discontinuities in the solar wind}

\author[0000-0002-7597-4980]{T. Neukirch}
\affiliation{School of Mathematics and Statistics, University of St Andrews, St Andrews KY16 9SS, UK}       \author[0000-0002-4974-4786]{I.Y. Vasko}
\affiliation{Space Sciences Laboratory, University of California, Berkeley, CA 94720, USA}
\affiliation{Space Research Institute of Russian Academy of Sciences, Moscow, Russia}
\author[0000-0001-8823-4474]{A. V. Artemyev} 
\affiliation{Institute of Geophysics and Planetary Sciences, University of California, Los Angeles, USA} 
\affiliation{Space Research Institute of Russian Academy of Sciences, Moscow, Russia}
\author[0000-0003-2353-8586]{O. Allanson} 
\affiliation{Space and Atmospheric Electricity Group, Department of Meteorology, University of Reading, Reading, RG6 6BB, UK}                   
                   

\correspondingauthor{Thomas Neukirch}
\email{tn3@st-andrews.ac.uk}


\begin{abstract}
Kinetic-scale current sheets observed in the solar wind are frequently approximately force-free 
despite
the fact that their plasma $\beta$ is of the order of one. In-situ measurements have recently shown that plasma density and temperature often vary across the current sheets, while the plasma pressure is approximately uniform. In many cases these density and temperature
variations are asymmetric with respect to the center of the
current sheet.
To model these observations theoretically we develop
in this paper 
equilibria of kinetic-scale force-free current sheets that
have plasma density and 
temperature gradients. The models can also be useful for analysis of stability and dissipation of the current sheets in the solar wind.   
\end{abstract}



\section{Introduction}
\label{sec:introduction}

The early {\it in-situ} measurements in the solar wind indicated the ubiquity of magnetic field discontinuities or, equivalently, current sheets with spatial scales 
below a few tens of ion thermal gyroradii or ion inertial lengths \citep[e.g.,][]{Burlaga77,Tsurutani79,Lepping86}. The magnetic reconnection within these kinetic-scale 
structures may provide ion and electron heating \citep[e.g.,][]{Osman11,Gosling12,Pulupa14}, though the overall contribution of the current sheets to the solar wind 
heating is unknown \citep[e.g.,][]{Cranmer09}. The disruption of the kinetic-scale current sheets via the magnetic reconnection is potentially a mechanism resulting 
in the spectral break of the magnetic field turbulence spectrum at ion scales \citep[e.g.,][]{Mallet17,Franci17,Vech18}. The mechanisms responsible for formation of the current sheets 
include 
Alfven wave steepening \citep[e.g.,][]{Medvedev97:prl}
and the natural appearance of sheet-like structures in the course of development of the turbulence cascade \citep[e.g.,][]{Greco09,Greco16,Franci17}. 


The early {\it in-situ} measurements focused on classifying the current sheets in terms of tangential and rotational discontinuities based on the analysis of the 
magnetic field component perpendicular to the current sheet plane \citep[e.g.,][]{Tsurutani&Smith74,Burlaga77,Lepping86}. However, the estimates of the fraction 
of tangential and rotational discontinuities in the solar wind are still controversial \citep[e.g.,][]{Knetter04,Neugebauer06,Artemyev19:grl:solarwind}. The {\it in-situ} 
measurements unambiguously showed that current sheets in the solar wind are often approximately one-dimensional and force-free, i.e. the current density is 
mostly parallel to the magnetic field and the magnetic field rotates across a current sheet, while its magnitude remains constant 
\citep[e.g.,][]{Burlaga77,Lepping86,Neugebauer06,Paschmann13:angeo}. Recent statistical analyses \citep{Artemyev18:apj, Artemyev19:grl:solarwind} 
have shown that the plasma density $n$ and ion and electron temperatures $T_{i,e}$ typically vary across a current sheet. In these analyses it was also 
shown that the density and temperature variations are anti-correlated $\Delta T_{i,e}/T_{i,e}\propto -\Delta n/n$, so that the plasma pressure is essentially 
uniform across the current sheets as required by the pressure balance. 

Within the large number of known one-dimensional kinetic current sheet models \citep[e.g.,][]{Lemaire&Burlaga76,Bobrova79,Roth96,Kocharovsky10,Panov11:magnetopause}, the most relevant to 

the 
solar wind observations mentioned above
are the recently 
developed models of force-free current sheets representing tangential \citep{Harrison09:prl,Wilson&Neukirch11,Allanson15} and rotational \citep{Artemyev11:pop,Vasko14:angeo_by} discontinuities. In these kinetic models of both force-free tangential and rotational discontinuities the plasma density and temperature are uniform across the current sheet. 

We remark that there is a much broader class 
of collisionless tangential discontinuity models \citep{Roth96} that can in principle be used to describe magnetic fields of 
solar wind discontinuities \citep{deKeyser96, deKeyser97:conf} and does even allow for the inclusion of plasma velocity 
shear \citep{deKeyser97, deKeyser13}, which is observed for some solar wind discontinuities \citep{deKeyser98:grl, Paschmann13:angeo, Artemyev19:grl:solarwind}.

These models start from specifying the dependence of the distribution functions on the constants of motion and have been
developed to give a detailed description different plasma populations in magnetic current sheets \citep{Roth96}. 
When starting from specifying the particle distribution functions any self-consistent model of a collisionless configuration has to be completed by solving Maxwell's equations.
With the form of the distribution functions used for a detailed description of current sheets \citep[see e.g. the model-data comparison in][]{deKeyser96, deKeyser97} it is usually not possible to obtain analytical solutions 
for the electromagnetic fields and hence these have to be determined using numerical methods. This in turn implies that the exact spatial variation of the particle densities, the pressure and the temperature 
is only available after the numerical calculation of the electromagnetic fields has been carried out.

In this paper we use a different approach, mainly for two reasons. Firstly, as already mentioned above, the magnetic field configuration of many of the current sheets observed in the solar wind 
is observed to be force-free  to a good approximation \citep[e.g.,][]{Artemyev19:jgr:solarwind}. For one-dimensional tangential discontinuities this directly implies that the magnetic field strength $|\mathbf{B}|$ and the
plasma pressure do not vary across the discontinuity \citep[see e.g.][]{Harrison09, Neukirch18}. This puts additional constraints on the possible dependence of the particle distribution functions on the constants of motion
and makes finding such distribution functions for force-free magnetic field configurations non-trivial. As a number of self-consistent distribution functions for the force-free version of the Harris sheet \citep{Harris62} 
have been found \citep[e.g.][]{Harrison09:prl,NWH09,Wilson&Neukirch11,Kolotkov15, Allanson15,wilson17,Wilson18:cs}, we use one of those force-free distribution functions as a starting point for the investigation in this paper.

Moreover, secondly, in the case we consider in this paper the process of determining appropriate distribution functions for a force-free magnetic tangential discontinuity starts from a known electromagnetic field
configuration (here the force-free Harris sheet) and one determines compatible distribution functions that lead to a self-consistent equilibrium by solving this "inverse" problem
\citep[see e.g.][]{Allanson16:jpp,Allanson18,Neukirch18}. Starting from an analytically known magnetic field configuration and corresponding distribution functions as a starting point of the investigation 
allows us more control direct control. An additional advantage of a completely analytical approach could be that it usually simplifies the implementation of the kinetic equilibrium as initial conditions in numerical simulations using,
for example, particle-in-cell (PIC) codes.

The crucial point is that none of the currently known collisionless force-free current sheet models is capable of describing the recently observed non-uniform density and temperature profiles in the solar wind current sheets.
Thus, it is our main motivation to develop analytical kinetic models of force-free current sheets that include the observed features.
From a more theoretical point of view, the development of analytical kinetic current sheet models including the observed gradients
will also simplify further investigations of their dynamics. For example, it is known that the stability of current sheets is rather sensitive to the initial equilibrium 
configuration \citep[e.g.,][]{Pucci18}. 
Moreover, PIC simulations have recently shown that the nonlinear evolution of the reconnection process and particle 
acceleration 
is strongly dependent 
on the presence of the guide field and plasma density and temperature gradients across the current sheets \citep[e.g.,][]{Wilson16:cs,Lu19:apj}. 

In this paper we present observations of the solar wind current sheets with plasma density and temperature gradients and develop a class of 
collisionless force-free equilibrium models
that incorporate the observed (asymmetric) variations of the plasma density and temperature. 

\begin{figure*}
 \centering
    \includegraphics[width=0.6\textwidth]{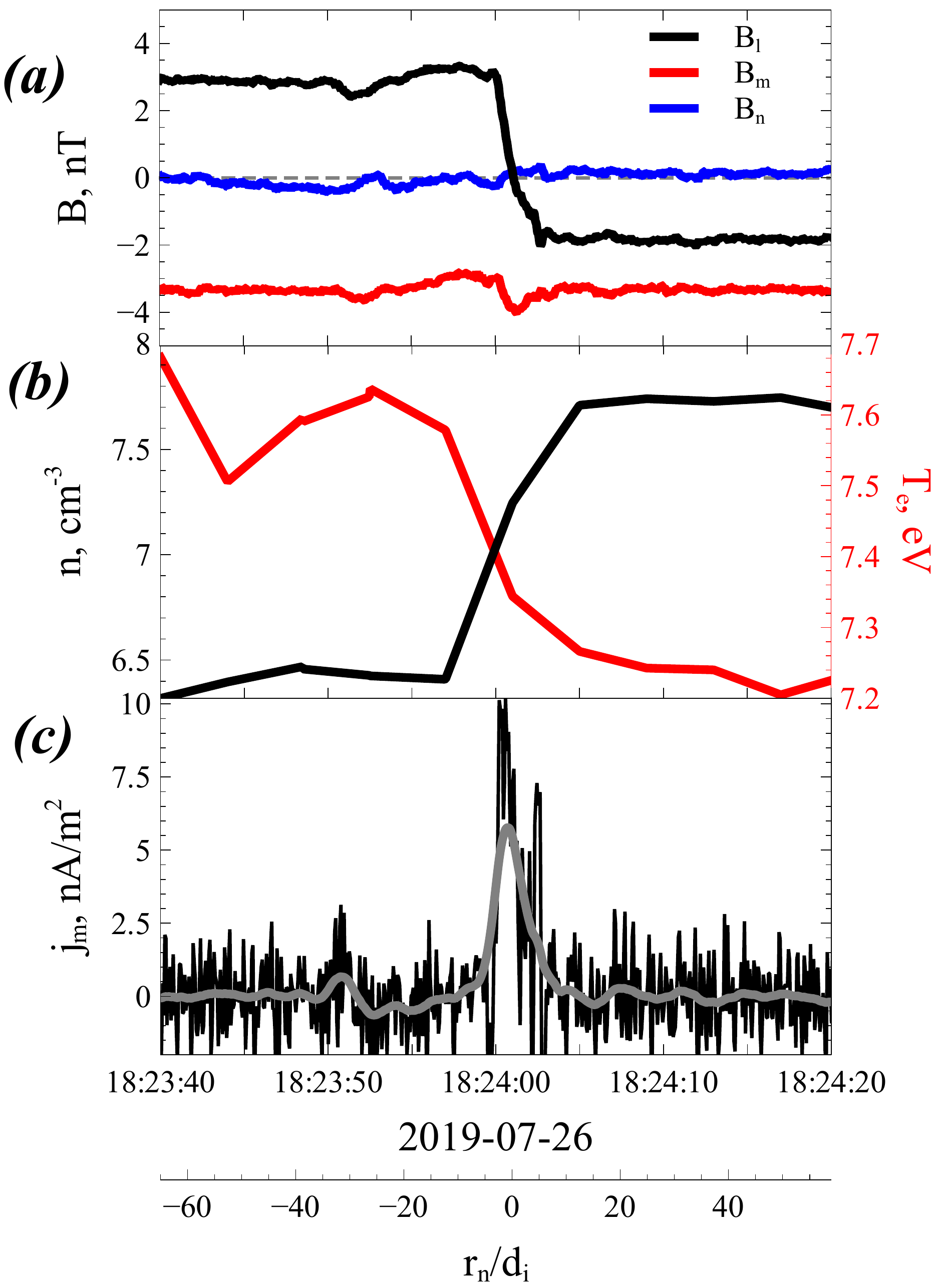}
  \caption{An example of a current sheet crossing by ARTEMIS spacecraft: (a) three magnetic field components in the local coordinate system \citep{Sonnerup68} with the additional constraint $\langle B_n\rangle = 0$ \citep[see section 8.2.6 in][]{bookISSI}, (b) electron density and temperature measurements, (c) current density profile (grey color shows smoothed profile). Bottom horizontal axis show spatial coordinate across the sheet (normalized on the ion inertial length, $d_i$).  \label{fig0}}
\end{figure*}

\begin{figure*}
 \centering
    \includegraphics[width=0.9\textwidth]{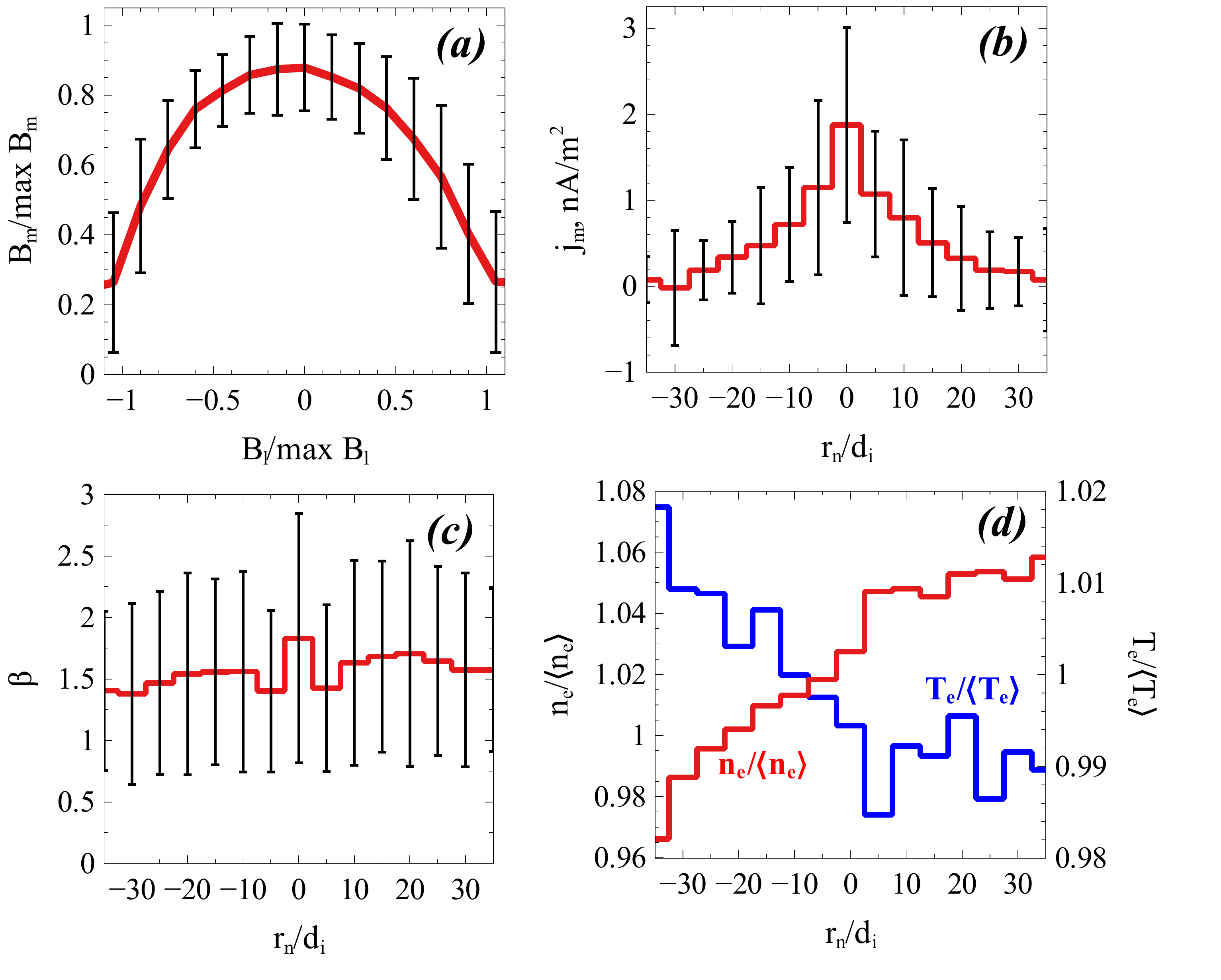}
  \caption{Average profiles of magnetic field, current density, plasma characteristics for a dataset of $\sim 200$ discontinuities observed by the ARTEMIS spacecraft in the near-Earth solar wind \citep[see details of the dataset in][]{Artemyev19:jgr:solarwind}. The main criterion of discontinuity selection to the dataset is the peak current density exceeding $1$ nA/m$^2$. Black error bars show the standard deviation. In each case, electron densities and temperatures are normalized by the average value across the discontinuity. Orientation of $r_n$ is chosen to have $dn_e/dr_n>0$ for all selected discontinuities. 
  \label{fig1}}
\end{figure*}


\section{Observations}
\label{sec:observations}

We present observations of current sheets 
by the
ARTEMIS spacecraft, which probes the solar wind at a few tens of Earth radii upstream of the Earth's bow shock \citep{Angelopoulos11:ARTEMIS}. We use the magnetic field measurements with temporal resolution of 5 vectors per second \citep{Auster08:THEMIS} and measurements of electron density and temperature available at 4s cadence 
(all plasma parameters are measured by electrostatic analyzers onboard ARTEMIS, see \citet{McFadden08:THEMIS}).

Figure \ref{fig0} presents an example of a particular current sheet observed aboard ARTEMIS. Panel (a) presents the magnetic field in the coordinate system ({\bf l},{\bf m},{\bf n}) determined using the Minimum Variance Analysis (MVA) \citep{Sonnerup68}. The magnetic field component $B_n$ is perpendicular to the current sheet plane, $B_{l}$ 
reverses the sign across the current sheet, $B_{m}$ is the so-called guide field. In a 1D approximation all 
variables vary across the current sheet that is along the normal ${\bf n}$. Panel (a) shows that the current 
sheet is approximately force-free, because $B_{l}^2+B_{m}^2\approx {\rm const}$. For the single spacecraft measurements the 
determination of the normal ${\bf n}$ is generally not sufficiently 
accurate \citep[e.g.,][]{Horbury01, Knetter04} to separate rotational and tangential discontinuities. Thus, we assume that the observed 
discontinuity is tangential and apply an additional constraint to the local coordinate system, namely $\langle B_n\rangle = 0$ \citep[see section 8.2.6 in][]{bookISSI}.
Panel (b) shows that the plasma density and electron temperature 
variations across the current sheet are anti-correlated. The plasma density increases across the current 
sheet by about 20$\%$, while the electron temperature decreases by about $5\%$. ARTEMIS measurements of 
the ion temperature in the solar wind are much less accurate than electron temperature measurements. The 
assumption of the pressure balance across the current sheet suggests that the ion temperature should also 
decrease across the current sheet by a few tens of percent. Because the Taylor hypothesis applies for the 
current sheets in the solar wind, we can estimate the current densities $j_{l}\propto -dB_m/dt$ and 
$j_{m}\propto dB_{l}/dt$ \citep[see][for details]{Artemyev19:jgr:solarwind}. Panel (c) shows that the 
current density reaches values of 10 nA/m$^{2}$, which is comparable to the highest current densities in 
the solar wind \citep[e.g.,][]{Podesta17}. The use of the Taylor hypothesis allows translating the 
observations in time into space. The spatial axis in Figure \ref{fig0} shows that the current sheet is 
an ion-scale structure with the thickness of a few ion inertial lengths or, equivalently, a few hundred 
kilometers. To demonstrate that the current sheet in Figure \ref{fig0} is not exceptional, we use a 
dataset of more than four hundred current sheets collected by the ARTEMIS spacecraft over two years of 
observations \citep[see ][for details]{Artemyev19:jgr:solarwind}.

Figure \ref{fig1} presents the averaged properties of the selected current sheets. Panel (a) shows the current sheets in the solar wind typically have a {\it half-ring} $B_{m}$ vs. $B_{l}$. This is equivalent to the statement that ion-scale current sheets in the solar wind are predominantly force-free, i.e. $B_{m}^2+B_{l}^2\approx {\rm const}$. Panel (b) shows that the $B_l$ reversal across the current sheet 
corresponds
to the current density $j_m\sim 1$ nA/m$^2$ localized within about ten ion inertial lengths. Panel (c) shows that the plasma beta, 
$\beta=8\pi p/B^2$, 
is typically about unity and does not vary across the current sheets in accordance with the force-free nature of the current sheets. Panel (d) shows that though $\beta$ is approximately uniform across the current sheets, there are clearly variations of the plasma density and electron temperature. Statistically, the plasma density varies by about 10$\%$, while the electron temperature varies by about 3$\%$ across the current sheet.

Although there are kinetic current sheet models that are sufficiently flexible to describe a large variety of tangential discontinuities \citep[see e.g. the review by][and references therein]{Roth96}, these models generally require a numerical solution of Maxwell's equations to achieve self-consistency. This complicates the matching of these models to the observations, in particular with regards to 
the additional constraints that have
to be satisfied by distribution functions for force-free collisionless current sheets.
Therefore we will start from a kinetic current sheet model that is completely analytical and already satisfies the force-free condition.
However, there are currently no simple analytical kinetic current sheet models 
which incorporate all of the observed features: (1) the force-free current sheet with spatial scales of a few ion inertial lengths and $\beta$ of the order of unity; (2) anti-correlated plasma density and temperature variations across the current sheet. In the next section we develop kinetic models for such current sheets assuming that they are tangential in nature, that is $B_{n}=0$.

We should mention that not all the discontinuities that were observed (and included in our statistics) are tangential, but that distinguishing observationally between 
tangential and rotational discontinuities is not a well resolved problem \citep[see discussion in][]{Neugebauer06}. The dataset presented in Fig.\ref{fig1} has been collected by the two ARTEMIS probes, 
whereas at least four-spacecraft observations are required for an accurate determination of the local coordinate system and estimation of $B_n$ \citep{Knetter04}. Therefore, in this paper 
we focus on modelling tangential discontinuities and leave the question of the relative percentage of tangential versus rotational discontinuities within the total amount of solar wind discontinuities to future investigations.

Independently of the classification of the discontinuities, the observations of these plasma structures in the solar wind are often associated with measurements of plasma shear flow 
\citep{deKeyser98:grl, Paschmann13:angeo, Artemyev19:jgr:solarwind}. This shear flow, which is related to the cross-field plasma (both ion and electron) velocity, 
can result in 
the generation of polarization electric fields \citep[e.g.,][]{Roth96, deKeyser13} that are enhanced by plasma pressure gradients across the discontinuities \citep[e.g.,][]{YL04, Lu19:jgr:cs}. 
However, there are no such gradients in force-free discontinuities. Moreover, some population of these discontinuities have the main magnetic field reversal 
along solar wind flow, i.e. the plasma shear flow is along the magnetic field and there is almost no cross-field shear flow. 
For this type of discontinuity the effect of the polarization electric field is negligible. In this paper we will focus on the theoretical description of this type of discontinuity 
and will not consider a finite electric field. A more general case could, for example, be described in future studies following the approach from \citet{deKeyser13}.

\section{Kinetic model of a force-free tangential discontinuity}
\label{sec:kinmodel}

In this section the local coordinate system $({\bf l},{\bf m},{\bf n})$ is denoted $(x,y,z)$. We consider a 
one-dimensional current sheet with the magnetic field ${\bf B}=B_{x}(z) {\bf e}_{x}+B_y(z) {\bf e}_y$. 
The development of a stationary kinetic current sheet model requires to provide a class of electron 
and ion distribution functions $F_{i,e}({\bf v},z)$, which would result in the current density 
${\bf j}=j_{x}(z){\bf e}_{x}+j_y(z){\bf e}_y$ consistent with the magnetic field ${\bf B}$, and the desired spatial 
distribution of the plasma density and ion and electron temperatures across the current sheet. The particle distribution 
functions, being solutions of the Vlasov equation, can be written as functions of the integrals of particle motion 
\citep[e.g.,][]{Schindler10}. In the considered one-dimensional current sheet there are three integrals of particle motion: 
the total energy $H_{s}=m_{s} {\bf v}^2/2+q_{s}\Phi$ and generalized momenta
$p_{xs}=m_{s}v_{x}+q_{s}A_{x}/c$ and $p_{ys}=m_{s}v_{y}+q_{s}A_{y}/c$, 
where $\Phi(z)$ is the electrostatic potential, ${\bf A}=A_{x}(z){\bf e}_{x}+A_y(z){\bf e}_y$ is the vector potential, 
$m_{s}$ and $q_{s}$ are particle mass and charge, $s=i,e$ correspond to ions and electrons ($q_{i}=-q_{e}\equiv e$).

\begin{figure*}
   \centering
   \vspace{5cm}
   \includegraphics{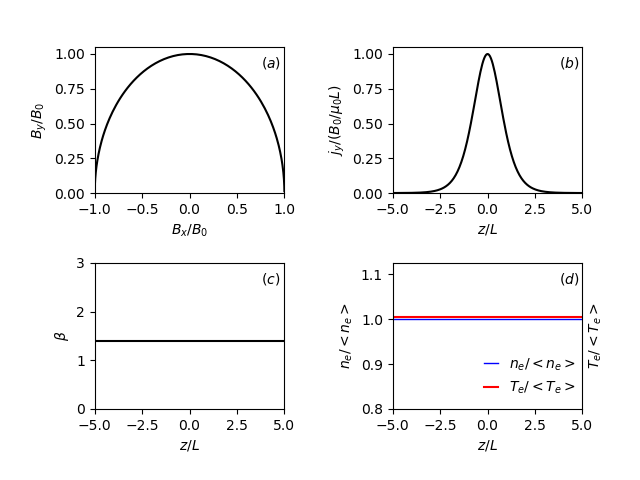} 
   \caption{The summary of the macroscopic properties of the current sheet models developed by \cite{Harrison09}: (a) the  half-ring $B_{x}$ vs. $B_{y}$ shape is due to $B_x^2+B_y^2=B_0^2$; (b) the profile of the $y$-component of the current density (c,d) the plasma $\beta$ is generally above unity, while the plasma density and electron and ion temperatures are uniform across the current sheet.}
   \label{fig:ffh}
\end{figure*}

Figure \ref{fig:ffh} illustrates the macroscopic quantities consistent with the class of kinetic models of 
force-free current sheets with the magnetic field $B_x(z) = B_0 \tanh(z/L)$ and $B_y(z) = B_0\cosh^{-1}(z/L)$ developed by 
\cite{Harrison09:prl}. In that class of models $B_{x}^2+B_y^2=B_0^2$, the plasma density and particle temperatures 
are uniform across the current sheet, and the model parameters are chosen in such a way that the 
electrostatic field vanishes identically, $\Phi(z)=0$. The plasma $\beta$ is above 
unity in the original class of \cite{Harrison09:prl} models, but $\beta$ can be arbitrary in more generalized 
models \citep[see][for a review]{Neukirch18}. The simplest example from that class of models is the one with the 
ion distribution function given by the Maxwellian distribution, $F_{0i}({\bf v},z)=n_0 (m_i/2\pi T_i)^{3/2}\exp(-H_i/T_i)$, and electron distribution function given as follows
\begin{eqnarray*}
    &F_{0e}&({\bf v},z)=n_{0} (1+b)^{-1}\left(\frac{m_{e}}{2\pi T_{e}}\right)^{3/2}\exp(-H_{e}/T_{e})\cdot\\
    &\cdot&[b+\exp(u_{0}p_{ye}/T_{e})\exp(-m_{e}u_0^2/2T_{e})-\\&-&\frac{1}{2}\cos(u_{0}p_{xe}/T_{e})\exp(m_{e}u_0^2/2T_{e})],
    \label{eq:electronDF}
\end{eqnarray*}
where $n_0$ is the plasma density, $T_{e}$ and $T_{i}$ are electron and ion temperatures (here and in the remainder of this paper we absorb the Boltzmann constant factor, $k_B$ into the temperature), 
$u_0$ is related to $B_0$ and $L$ by the relations 
$en_0u_0=-cB_0/4\pi L$
and 
$B_0 L = - 2 T_e/e u_0$. 
This implies that $L^2 = c T_e/2 \pi e^2 n_0 u_0^2$.
The electron temperature determines the 
amplitude of the magnetic field, 
$B_0^2=8\pi n_0 T_{e}$. 
The parameter $b$ sets the density of the background 
electron population not contributing to the current density, it has to be large enough to keep the electron velocity distribution function positive. 
In what follows we generalize the models developed by \cite{Harrison09:prl} to have the asymmetric distribution of the plasma density across the current sheet similar to that in Figures \ref{fig0} and \ref{fig1}.

The models of force-free current sheets with asymmetric plasma density profile can be developed within rather wide class of particle velocity distribution functions: $F_{s}=F_{0s}(H_{s},p_{xs},p_{ys})+\Delta F_{s}(H_s,p_{xs})$,
where $F_{0s}$ is, for example, the class of distribution functions suggested by \cite{Harrison09}, while $\Delta F_{s}(H_s,p_{xs})$ corresponds to additional electron and ion populations. 
In principle, $F_{0s}$ can be any distribution function consistent with the magnetic field profile \citep[e.g.][]{Kolotkov15,Allanson15,Allanson16:jpp,wilson17, Wilson18:cs}. The distribution function of the additional populations should be chosen so that they provide no contribution to the current density, $\int v_{x}\Delta F_{s} d^{3}{\bf v}=0$, but contribute to the density $\int \Delta F_{s}(H_s,p_{xs}) d^3{\bf v}=\Delta n_{s}(A_x)$, where $\Delta n_{s}$ should be an odd 
function of $A_{x}$ that is $\Delta n_{s}(-A_x)=-\Delta n_{s}(A_x)$. In that case the magnetic field remains identical to that in the models of \cite{Harrison09:prl}, while the electron density distribution will be asymmetric across the current sheet, because $A_x(z) = B_0 L\arctan[\sinh(z/L)]$ is asymmetric with respect to $z=0$. Because the magnetic field configuration remains force-free that is $B^2={\rm const}$, the pressure balance across the CS 
$p_{zz}+B^2/8\pi={\rm const}$ 
results in a constant $zz-$component of the pressure tensor, $p_{zz}={\rm const}$. For a non-uniform plasma density the variation of the temperature 
$T_{zz}$ 
across the current sheet is anti-correlated with the density variation. 

One of the simplest choices of the velocity distribution functions of the additional populations is $\Delta F_{s}=g_{s}(H_{s}) u_0 p_{xs}/T_{e}$, where $g_{s}(H_{s})$ should satisfy $\int v_{x}^2 g_{s}(H_{s}) d^3{\bf v}=0$. The class of functions $g_{s}(H_{s})$ satisfying the latter condition is rather broad, while a particular example is 
\begin{eqnarray*}
g_{s}&=&\delta n_{s} \left(\frac{m_{s}}{2\pi T_{s}}\right)^{3/2}\frac{\kappa_{s}^{5/2}e^{-\kappa_{s}H_{s}/T_{s}}-\tilde{\kappa}_{s}^{5/2}e^{-\tilde{\kappa}_{s}H_s/T_{s}}}{\kappa_{s}-\tilde{\kappa}_{s}}
\end{eqnarray*}
where $\delta n_{s}$, $\kappa_{s}$ and $\tilde{\kappa}_{s}$ are free parameters. The number of free parameters can be reduced by taking the
limit $\tilde{\kappa}_{s} \to \kappa_{s}$, which leads to the following class of $\Delta F_{s}$
\begin{eqnarray*}
\Delta F_s & =& \delta n_{s} \left(\frac{\kappa_s m_{s}}{2\pi T_{s}}\right)^{3/2}\left(\frac{5}{2}-
    \frac{\kappa_s H_s}{T_{s}} \right)
    e^{-\kappa_s H_s/T_{s}}\frac{u_0 p_{xs}}{T_{s}},
    \label{eq:deltaF}
\end{eqnarray*}

The additional particle density is given by
\begin{equation}
 \Delta n_s = -\frac{q_s}{e} \delta n_{s} \exp( -q_s \kappa_s \beta_s \Phi)(1 -q_s \kappa_s \beta_s \Phi)
     \frac{2  A_x}{B_0 L} .
     \label{eq:deltan_s_v2}
\end{equation}
The quasi-neutrality condition $\sum_s q_s n_s =0$ has the solution $\Phi=0$, if we let
$\delta n_{i} = - \delta n_{e}$.

The expression in Eq.\ (\ref{eq:deltaF}) is a relatively
simple member of a wider class of functions with the desired property that they contribute to the particle density, 
but not to the current density (if $\Phi=0$). We remark that by choosing parameters appropriately it is always possible to ensure that the total DF, $F_s + \Delta F_s$, is positive definite. 

With $\Phi=0$, $q_e =-e$, $q_i=e$ (and defining 
$\delta n_{e} = \epsilon n_0$) the additional density term is given by 
\begin{equation}
    \Delta n_s  = \Delta n= \epsilon n_0 
    \frac{2 A_x}{B_0 L}.
    \label{eq:deltan}
\end{equation}

Because $A_x(z)$ is asymmetric with respect to $z$, $\Delta n_s$ introduces the desired density asymmetry across the current sheet. 
If we define the temperature via the equation $p_{zz, e} = k_{B} T_e n_e$, the temperature will also be asymmetric due to the pressure remaining constant,
as found in the observations \citep{Artemyev19:grl:solarwind}.

\begin{figure*}
 \centering
    \includegraphics[width=0.9\textwidth]{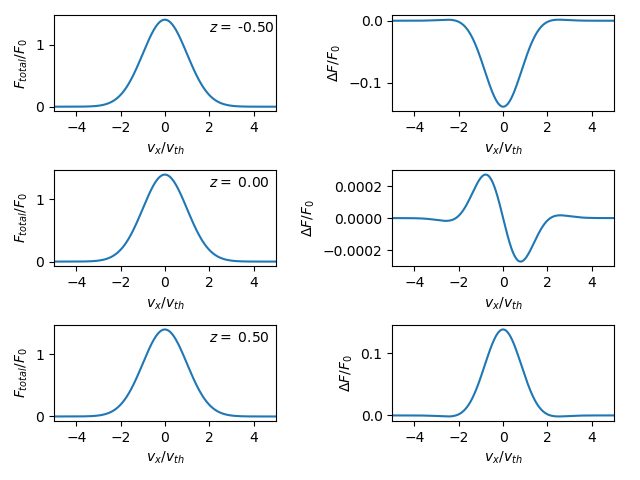}
    \caption{In the left column the dependence of the total electron DF $F_{0e} + \Delta F_e$ on $v_x$ (for $v_y = v_z =0$) is shown at three different positions: $z/L = -0.5$ (top row), $z/L = 0.0$ (middle row) and 
    $z/L = 0.5$ (bottom row). The right column shows the same plots for $\Delta F_e$ alone. Here $\epsilon = 0.05$ and $u_0/v_{th,e} = -3.9\cdot 10^{-3}$.}
\label{fig:exampleDFe}
\end{figure*}

\begin{figure*}
 \centering
    \includegraphics[width=0.9\textwidth]{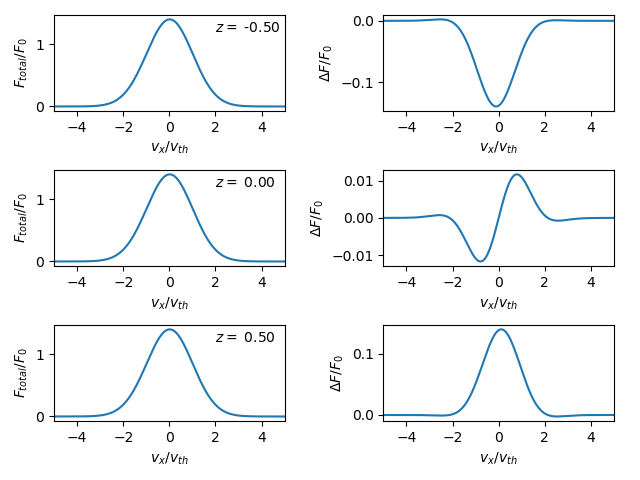}
    \caption{The same plots as in Fig. \ref{fig:exampleDFe}, but for the ions. The only noticeable difference to the plots for the electrons is the larger amplitude of $\Delta F_i$ at $z=0$ and the symmetry reversal of its minimum and maximum values.}
\label{fig:exampleDFi}    
\end{figure*}

In order to construct a realistic example we now assume that $L/d_i = 10$ (the ratio of the current sheet width to the ion inertial length), $\beta_{p} = 1.4$, $T_e/T_i = 1.0$ and $m_i/m_e =1836$. We then find
that $u_0/v_{th,e} =  \sqrt{2 m_e/m_i \beta_p} d_i/L \approx - 3.9 \cdot 10^{-3}$. Using $\kappa_e = \kappa_i = 1.1$ and $\epsilon = 0.05$, both 
distribution functions can be shown to be positive. We show example plots of the variation of the full electron and the ion distribution
 functions with $v_x$ (for fixed values of $z$, $v_y$ and $v_z$) in Figs.\ \ref{fig:exampleDFe} and \ref{fig:exampleDFi}.
Due to the relatively small value of $u_0/v_{th,e}$, the difference between the total electron and ion distribution functions is also very small. In the same figures we also show how $\Delta F_s$ varies with $v_x$.
    
\begin{figure}[ht!]
    \centering
   \includegraphics[width=0.45\textwidth]{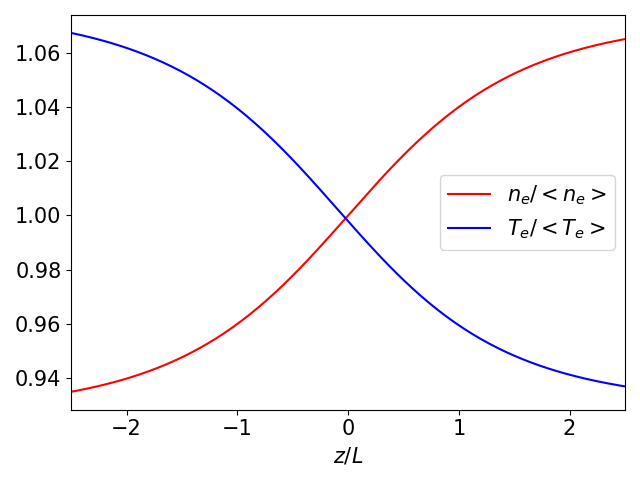}
    \caption{An example of asymmetric density and temperature profiles resulting from the theoretical models ($\epsilon = 0.05$).}
    \label{fig:exampledensity}
\end{figure}

In Fig. \ref{fig:exampledensity} we show the resulting modified density and temperature profiles for the same parameter values that were used for the distribution function plots. As desired the density and temperature profiles show the general behaviour that is also seen in the observation shown in Fig.\ \ref{fig1}

The structure of the distribution function in velocity space is seen to be very close to a Maxwellian distribution function (see Fig. \ref{fig:exampleDFe} and Fig. \ref{fig:exampleDFi}). This structure suggests that the distribution functions presented in this paper are likely to be stable to small perturbations \citep[e.g. see standard stability arguments by][]{Gardner-1963, bookKrall&Trivelpiece73}.

\section{Discussion and Conclusion}

Recent spacecraft observations have shown that current sheets in the solar wind frequently exhibit non-symmetric and anti-correlated electron density and temperature distributions with respect to the current sheet center \citep{Artemyev18:apj}. The origin and effects of these features on the stability of the current sheets in the solar wind remains unknown, partly due to absence of kinetic models that could be used in the stability analysis. 
Self-consistent kinetic models of force-free current sheets have only been developed relatively recently \citep[e.g][]{Harrison09:prl} and in these models the plasma and temperature distributions are uniform.

In this paper we have demonstrated that by adding a suitable further term to the distribution 
function of \citet{Harrison09:prl} it is possible to generate self-consistent kinetic equilibria which
have asymmetric spatial profiles of particle density and temperature, while retaining the macroscopic current sheet
equilibrium unchanged. We have presented an illustrative example which showed that for parameter values which are 
typical for solar wind current sheets observed at 1 A.U., one can easily find self-consistent particle distributions functions giving
rise to macroscopic spatial variations in particle density and temperature that closely resemble those found in the observations.

The work presented in this paper could be further extended in a number of ways. For example, instead of using the
distribution functions of \citet{Harrison09:prl} as $F_{0s}$, one can in principle choose any other particle distribution function giving rise to the same magnetic field profile. While the distribution functions used for $F_{0s}$ in this paper always lead
to an equilibrium with plasma $\beta >1$ as well as spatially constant density and temperature profiles, other distribution functions allow for values of plasma $\beta <1$ \citep[e.g.][]{Allanson15, Allanson16:jpp,Wilson18:cs} or for additional symmetric variations in the particle density and temperature \citep[e.g.][]{Kolotkov15}.

Another possible extension of the work presented here relates to the specific and relatively 
simple form for $\Delta F_s$ that we have used. This form for $\Delta F_s$ is just one example taken from a family of
possible $\Delta F_s$; other examples include $\Delta F_s \propto \sin(K_s p_{xs}) g_s(H_s)$ and 
$\Delta F_s \propto \exp(K_s p_{xs}) g_s(H_s)$ (with $\int v_{x}^2 g_{s}(H_{s}) d^3{\bf v}=0$ and $K_s$ a model dependent constant).

It is also important to point out that within the same class of particle velocity distribution functions 
one can develop models of force-free current sheets with symmetric 
density profiles having either maximum or minimum in center of the current sheet (similar to models of \cite{Kolotkov15}). 
The symmetric profiles of the plasma density are obtained for 
distribution functions 
for which
$\Delta n_{s}(A_x)$ is an even function of $A_x$. 
The additional population should not contribute to the current density and the simplest choice of such particle 
distribution functions is $\Delta F_{s}=g_{s}(H_s) (\beta_e u_0 p_{xs})^2$, where $g_{s}(H_{s})$ should again 
satisfy the condition $\int v_{x}^2 g_{s}(H_{s})d^{3}{\bf v}=0$. For the example distribution function given above,
using the same $g_s(H_s)$ that was used in section \ref{sec:kinmodel} the plasma 
density is as 
$n_{s}=n_0(1 + \epsilon 4 A_x^2/B_0^2L^2)$.
As $A_x(z)$ is an odd function of $z$, $A^2_x(z)$ is
an even function of $z$ and the density profile is symmetric 
with respect to the current sheet center.

The self-consistent kinetic current sheet models presented here could, for example, be used as initial conditions for future analyses of collisionless kinetic processes involving tangential discontinuities in the solar wind plasma.


\acknowledgments
The authors would like to thank the referee for very useful and constructive comments that helped to improve the paper.

TN acknowledges financial support by the UK's Science and Technology Facilities Council (STFC) via Consolidated Grant  ST/S000402/1. OA was supported by the Natural Environment Research Council (NERC) Highlight Topic Grant \#NE/P017274/1 (Rad-Sat).

ARTEMIS data analysis was supported by NASA Grant NNX16AF84G and NASA contract NAS5-02099. We would like to thank the following people specifically: C.W. Carlson and J. P. McFadden for use of ESA data; K. H. Glassmeier, U. Auster, and W. Baumjohann for the use of FGM data provided under the lead of the Technical University of Braunschweig; and with financial support through the German Ministry for Economy and Technology and the German Aerospace Center (DLR) under contract 50 OC 0302. THEMIS data (ESA and FGM) were obtained from http://themis.ssl.
berkeley.edu/. Data access and processing were done using SPEDAS V3.1; see \citet{Angelopoulos19}.

\end{document}